\documentclass{nature}
\usepackage{xcolor}
\usepackage{blindtext}
\usepackage{caption}%add by bch
\usepackage{amsmath}
\usepackage{bm}
\usepackage{braket}
\usepackage{amsmath}
\usepackage{booktabs}%
\usepackage{amssymb}
\usepackage{textcomp}
\title{Origin of the insulating and superconducting phases in molecular solid $\kappa$-(BEDT-TTF)$_2$Cu$_2$(CN)$_3$}

\author{Dongbin~Shin$^{1,2\dagger}$, Fabijan Pavo\v{s}evi\'{c}$^3$, Nicolas Tancogne-Dejean$^{2}$, Michele Buzzi$^{2}$, Emil Vi\~nas Bostr\"om$^{2}$, Angel Rubio$^{2,4,5\dagger}$}
\usepackage{graphicx}
\makeatletter
\let\saved@includegraphics\includegraphics
\AtBeginDocument{\let\includegraphics\saved@includegraphics}
\makeatother
\begin{document}
%\linenumbers
\maketitle

\begin{affiliations}
 \item Department of Physics and Photon Science, Gwangju Institute of Science and Technology (GIST), Gwangju 61005, Republic of Korea 
 \item Max Planck Institute for the Structure and Dynamics of Matter and Center for Free Electron Laser Science, 22761 Hamburg, Germany
 \item Algorithmiq Ltd., Kanavakatu 3C, FI-00160 Helsinki, Finland
 \item Nano-Bio Spectroscopy Group, Departamento de Fisica de Materiales, Universidad del País Vasco UPV/EHU- 20018 San Sebastián, Spain
 \item Center for Computational Quantum Physics (CCQ), The Flatiron Institute, 162 Fifth avenue, New York NY 10010, USA
\end{affiliations}

\begin{abstract} 
Recent studies of organic molecular solids are highlighted by their complex phase diagram and light-induced phenomena, such as Mott insulator, spin liquid phase, and superconductivity.
However, a discrepancy between experimental observation and first-principle calculation on the $\kappa$-(BEDT-TTF)$_2$X family inhibits understanding their properties.
Here, we revisit the electronic structure of $\kappa$-(BEDT-TTF)$_2$Cu$_2$(CN)$_3$ with the recently developed DFT+GOU method to correct the energy level of molecular orbital states in the molecular solid.
Our work reveals that the insulating electronic structure of $\kappa$-(BEDT-TTF)$_2$Cu$_2$(CN)$_3$ originates from the energy gap between the highest occupied and the lowest unoccupied molecular orbital states of the BEDT-TTF dimers, that are the periodic unit of the molecular solid.
We verify that our calculation result provides consistent band gap, optical conductivity, and evolution of the metal-insulator transition as a function of pressure with experimental observations.
Especially, the superconducting dome of $\kappa$-(BEDT-TTF)$_2$Cu$_2$(CN)$_3$, which originates from the flat band state at the Fermi level, is reproduced. 
Additionally, we constructed a new low-energy lattice model based on the ability of electronic structure data that can be used to address many-body physics, such as quantum spin liquid and double-holon dynamics. 
Our provides a deeper understanding of the complex phase diagram and various light-induced phenomena in the $\kappa$-(BEDT-TTF)$_2$X family and the other complex organic molecular solids. 
\end{abstract}

\clearpage

\renewcommand{\thetable}{\textbf{Table \arabic{table} $\bm{|}$}}
\renewcommand{\thefigure}{\textbf{Fig. \arabic{figure} $\bm{|}$}}
\setcounter{figure}{0}

Molecular solids have attracted significant attention due to their complex phase diagrams and exceptional light-matter interactions.\cite{kurosaki_mott_2005,buzzi_phase_2021,buzzi_photomolecular_2020,mitrano_possible_2016,rowe_resonant_2023,zadik_optimized_2015,isono_quantum_2016,komatsu_realization_1996}
For instance, experimental observations reveal that $\kappa$-(BEDT-TTF)$_2$Cu$_2$(CN)$_3$ ($\kappa$-ET-CN) exhibits antiferromagnetic insulating behavior under ambient conditions but shows metal-insulator transition under pressure, with critical pressure of $0.37$~GPa, including a superconducting phase.\cite{lefebvre_mott_2000,limelette_mott_2003,komatsu_realization_1996,kurosaki_mott_2005}
Additionally, $\kappa$-ET-CN is a potential candidate for the quantum spin liquid phase, characterized by long-range spin entanglement and frustrated spin ordering due to quantum fluctuations.\cite{pratt_magnetic_2011,isono_quantum_2016,miksch_gapped_2021,suzuki_mott-driven_2022}
A similar phase diagram is also observed in the other organic molecular solids, such as the alkali-doped C$_{60}$ family, which shows antiferromagnetic insulating and superconducting phases depending on the external pressure.\cite{zadik_optimized_2015,nomura_unified_2015}
On the other hand, advanced theoretical approaches are needed to describe the properties of molecular solids rigorously.
For example, dynamical mean-field theory (DMFT) calculations based on the density functional theory (DFT) have been used to reproduce the superconducting phase diagram of alkali-doped C$_{60}$ under external pressure.\cite{nomura_unified_2015}
The tight-binding model Hamiltonian used as a starting point for the many-body correlated has also been applied for the $\kappa$-(BEDT-TTF)$_2$X ($\kappa$-ET family),\cite{Menke_Klett_Kanoda_Georges_Ferrero_Schäfer_2024,kandpal_revision_2009} but the metal-insulator transition under the external pressure remains not fully understood from the first-principle calculation.
These findings suggest that advanced theoretical work is necessary to elucidate the microscopic mechanism behind these intriguing phenomena in the molecular solids.\cite{buzzi_photomolecular_2020,mitrano_possible_2016,rowe_resonant_2023,warawa_ultrafast_2023,kawakami_nonlinear_2018}
Furthermore, recent experiments have reported light-enhanced superconductivity in $\kappa$-ET family\cite{buzzi_photomolecular_2020} as well as light-driven exotic charge dynamics in the $\kappa$-ET family, including ultrafast gap dynamics and nonlinear charge oscillations involving localized excited carriers on the molecular orbitals (MO).\cite{warawa_ultrafast_2023,kawakami_nonlinear_2018}

Recent developments in the DFT's description of exchange and correlation effects have enhanced the ability to describe the accurate electronic structure of condensed matter systems.\cite{Anisimov1991,heyd_hybrid_2003,onida_electronic_2002,beaulieu2021ultrafast} 
Conventional DFT employs a density functional exchange-correlation potential to approximate many-body effects within a mean-field framework, leading to successful computational studies in condensed matter physics and materials science.\cite{jones_density_2015}
However, DFT sometimes fails to accurately evaluate the electronic structure of correlated systems, such as metal-oxides, charge density wave states, open shell magnetic molecules, and organic molecular solids.\cite{Anisimov1991,Shin2021,nomura_unified_2015,cha_inaccuracy_2009}
This failure usually originates from the delocalized nature of the adiabatic density functional approximation employed to describe exchange-correlation interactions, where the electron-electron exchange is represented in terms of charge density rather than states interactions.\cite{andrade_prediction_2011,Janak1978,Shin2016}
Extended approaches to DFT calculations have been developed to address this limitation, such as DFT plus Hubbard U (DFT+U), hybrid functionals, and DFT+DMFT.\cite {Anisimov1991,Cococcioni2005,heyd_hybrid_2003,onida_electronic_2002,Agapito2015,Tancogne-Dejean2018,mohan_comparison_2010}
For instance, the hybrid functional, which employs the mixed description for exchange term between DFT functional and Hartree-Fock (HF), provides consistent band gap in solid and energy level of molecular systems as experimental observations under the screened Coulomb potential schemes.\cite{heyd_hybrid_2003,kronik_excitation_2012}.
In addition, the DFT+U method reproduces the Mott-insulating electronic structure in NiO by applying a U potential to the Ni $d$ orbitals.\cite{Anisimov1991} 
The inter-atomic site interaction correction (V) has recently been introduced in the DFT+U+V method, which can provide accurate optical responses in metal oxides.\cite{tancogne-dejean_parameter-free_2020,campo_extended_2010,lee_first-principles_2020}
The DFT+U with generalized orbital basis (DFT+GOU) method is developed to describe the insulating nature of 1T-TaS$_2$ by applying U potential to the charge density wave states.\cite{Shin2021}
This method especially enables correcting the on-site Coulomb interaction in generalized states, such as MOs and charge density wave states. 
In contrast, conventional DFT+U fails to address the on-site Coulomb interaction in these states, as it applies corrections only to atomic orbitals.\cite{Shin2021}
To reproduce the phase diagram of the molecular solid alkali-doped C$_{60}$ including its superconducting state, on the other hand, DFT calculation with the DMFT approach for electron and phonon systems is employed by considering the correlation of electron-electron and electron-phonon interactions.\cite{nomura_unified_2015}

Here, we investigate the insulating nature of the $\kappa$-ET-CN system using the recently developed DFT+GOU method.\cite{Shin2021} 
The $\kappa$-ET family shows discrepancies between experimental observations and DFT calculations on their phase diagram as a function of temperature and pressure.\cite{kurosaki_mott_2005,kandpal_revision_2009}
Although $\kappa$-ET-CN is experimentally observed as an insulator under ambient conditions,\cite{kurosaki_mott_2005} for example, static adiabatic DFT calculations produce a metallic band structure.\cite{kandpal_revision_2009}
This inconsistency inhibits the understanding of intriguing phenomena in the $\kappa$-ET family, such as metal-insulator transitions and light-induced superconductivity.\cite{buzzi_phase_2021,warawa_ultrafast_2023,kawakami_nonlinear_2018}
To resolve this problem, we first review how the band structure forms from the MO state in the simplified molecular solid model. 
Based on this idea, we describe the band structure of the solid starting from the MO energy level of the BEDT-TTF dimer that constitutes the periodic unit cell of the underlying molecular solid. 
By exploring the electronic structure with various DFT functionals, we verify that the corrected HOMO-LUMO gap of the BEDT-TTF dimer with the DFT+GOU method provides the insulating electronic structure of $\kappa$-ET-CN and its properties as experimental observations.
On the other hand, both the DFT and DFT+U methods fail to describe insulating electronic structure by underestimating the HOMO-LUMO gap of the parent dimer of a molecule.
In addition, varying external pressures reproduce the superconducting dome, starting from the insulating electronic structure in ambient conditions. 
We present a new set of tight-binding parameters for the insulating electronic structure of $\kappa$-ET-CN, which can be used for further studies on spin liquid and superconducting phases. 
Unlike previous models, our tight-binding model for $\kappa$-ET-X systems is constructed from the HOMO-LUMO gap of the dimer.
Our study provides insight into the insulating and superconducting nature of molecular solid systems based on the underlying molecule's electronic structure and vibrational properties.

\section*{Results}

Before establishing our main study on $\kappa$-ET-CN, we first introduce the most straightforward molecular solid system and its electronic structure obtained from the MO states.
One of the representative molecular solids is a half-filled 1D hydrogen chain model with a half-filled metallic single-band structure. 
When each of the two hydrogen atoms is dimerized, a gapped band structure emerges from the single metallic band due to dimerization between the two sites.
This phenomenon is known as Peierl's distortion.
From an equally spaced molecular distribution in the 1D chain structure (see Fig. 1(a)) with quarter-filling, a metallic band structure is achieved (see Fig. 1(b)). 
In this configuration, dimerization between two molecules (see Fig. 1(c)) can open a band gap by orbital hybridization, analogous to the previous case.
Here, we introduce a HOMO-LUMO gap ($\Delta$) of this dimer bonding orbital state ($\sigma$) for the anti-ferromagnetic solution.
It splits the $\sigma$ bonding states depending on the occupations for a single dimer, as depicted in Fig. 1(d). 
When these molecular states form a solid, the inter-site hopping ($t$) with the HOMO-LUMO gap ($\Delta$) in a single dimer determines the overall band structure and energy gap (see Fig. 1(e)). 
This example demonstrates that the basis molecule's MO-level structure is crucial for describing the electronic structure of molecular solids.

Formation of quasi-2D crystal consisting of two BEDT-TTF dimer sites in $\kappa$-ET-CN leads to the band dispersion from MO levels.
In the rhombohedral lattice, $\kappa$-ET-CN consists of two BEDT-TTF dimers forming a quasi-2D crystal and a Cu$_2$(CN)$_3$ metal frame layer divider with the space group $P21/c$, as shown in Fig.~2.
A quasi-2D rectangular lattice is constructed by two dimers (2(BEDT-TTF)$_2$) in the $b-c$ plane, as shown in Fig.~2(e).
By the charge transfer between (BEDT-TTF)$_2$ and Cu$_2$(CN)$_3$ layers, each (BEDT-TTF)$_2$ dimer becomes +1 cation and Cu$_2$(CN)$_3$ layer is -2 anion in $\kappa$-ET-CN system.
To understand the electronic structure of $\kappa$-ET-CN, we first focus on the MO levels in (BEDT-TTF)$_2^{+1}$ and 2(BEDT-TTF)$_2^{+1}$.
In the (BEDT-TTF)$_2^{+1}$, a small HOMO-LUMO gap is evaluated with spin doublet configuration (Fig.~2(d) and Table 1) with its odd number of electrons. 
In the 2(BEDT-TTF)$_2^{+1}$, inter-dimer hybridization in a vacuum leads to the smaller HOMO-LUMO gap compared to the single dimer case.
These conclusions are found to be robust to the level of theory and the choice of the exchange-correlation functional (see Tab. 1). 
We can expect a band dispersion for the $\kappa$-ET system from the HOMO-LUMO gap of the two-dimer system as a quarter-filled 1D chain case.
With Perdew-Burke-Ernzerhof (PBE) functional and van der Waals interaction,\cite{Perdew1996} we achieve the metallic band structure of $\kappa$-ET-CN (Fig.~2(g)), similar to the previous theoretical report, from a small HOMO-LUMO gap of two dimer systems.\cite{kandpal_revision_2009}
This result indicates that the $\Delta=0.11$~eV HOMO-LUMO gap (at the PBE+D3 level) induces a metallic band structure of $\kappa$-ET-CN.

The HOMO-LUMO gap of the charged two-dimer molecular system can be modulated by the U value in the DFT+GOU calculation by localizing more electrons on the MO states, thus correcting for the well-known failure of the local and semilocal DFT functionals. 
This behavior typically leads to an inaccurate evaluation of the energy level.
The conventional DFT calculation usually fails to describe the energy level of molecular systems compared to experimental observations.\cite{Shin2016,weigend_balanced_2005,mohan_comparison_2010}
To remedy this problem, we investigate the functional dependency on the HOMO-LUMO gap of 2(BEDT-TTF)$_2^{+1}$ using PBE, HF, PBE+GOU(ACBN0), and hybrid functional calculations as shown in Fig.~3(a).
As expected, the PBE functional predicts the lowest HOMO-LUMO gap ($\Delta^{DFT} = 0.11$eV), and the Hatree-Fock calculation provides the largest HOMO-LUMO gap ($\Delta^{HF} =2.4$eV) for a charged two-dimer system. 
This difference originates from the underestimated and overestimated on-site Coulomb interaction of DFT functional and HF calculation, respectively.\cite{Shin2016}
In addition, DFT+GOU(ACBN0)\cite{Agapito2015,tancogne-dejean_parameter-free_2020,Shin2021}, which determines the U value self-consistently as $U_{ACBN0}^{GOU}=2.1$~eV, provides a large HOMO-LUMO gap ($\Delta^{DFT+GOU}_{ACBN0}=2.0$~eV).
As the ACBN0 method is based on the HF-type Coulomb interaction term, it is reasonable that the corresponding DFT+GOU(ACBN0) provides a similar result as HF calculation.\cite{Agapito2015,Mosey2007,Mosey2008,tancogne-dejean_self-consistent_2017}
On the other hand, the hybrid functional (PW6B95) provides intermediate HOMO-LUMO gaps as $\Delta^{PW6B95}=0.66$eV.
To estimate the corresponding U value for each functional calculation, we evaluate the HOMO-LUMO gap with respect to the U parameter for DFT+GOU, as shown in Fig. 3(a).
This result indicates that there is a functional dependency on the HOMO-LUMO gap. 
Here, we show that a DFT+GOU with appropriate U values ($U \sim 0.3U_{ACBN0}^{GOU}$) can provide a consistent HOMO-LUMO gap with the result of the hybrid functional.
As a previous study reveals that the DFT+U fails to intentionally apply the Hubbard U potential to the MO or hybridized states,\cite{Shin2021} we verified that applying Hubbard U potential on C-$p$ orbitals and S-$p$ orbitals under the DFT+U scheme doesn't modify the HOMO-LUMO gap of the two-dimer system ($\Delta^{DFT+U} \sim 0.11$eV) significantly in the wide range of U values ($0\leq U\leq 10$~eV).
In SI, we discuss the accuracy of the DFT+GOU method depending on the subspace for the basis set and the effect of on-site Coulomb correction with an atomic orbital basis in DFT+U.

A large HOMO-LUMO gap of the charged two-dimer can lead to an insulating band structure for $\kappa$-ET-CN.
Given that hybrid functional calculations generally provide theoretical energy gaps in molecules and condensed matter systems that align closely with experimental observations, a hybrid functional approach would likely also accurately capture the electronic structure in our case.\cite{garza_predicting_2016,kronik_excitation_2012} 
However, the computational cost of such calculations is prohibitively high for the $\kappa$-ET-CN. 
As a practical alternative, we use the DFT+GOU method with an appropriate U value ($U\sim 0.24U_{ACBN0}^{GOU}=0.225$~eV) for MO of dimer in the $\kappa$-ET-CN ($U_{ACBN0}^{GOU}=0.94$~eV), which is achieved from the ratio between ACBN0 and HSE06 results for a HOMO-LUMO gap of two-dimer system ($U \sim 0.3U_{ACBN0}^{GOU}=0.63$~eV).
When we applied a Hubbard U ($U=0.225$~eV) on the MO states of a dimer ((BEDT-TTF)$_2$) in the $\kappa$-ET-CN, the insulating band structure with anti-ferromagnetic ordering configuration is achieved, as shown in Fig.~3(b).
This result indicates that the band gap ($E_{gap}$) of $\kappa$-ET-CN is well determined by the enough HOMO-LUMO gap ($\Delta$) of the dimers, similar to the 1D molecular solid example.
In addition, we scan the band gap of $\kappa$-ET-CN concerning the various U values for the dimer's MO states, as shown in Fig.~3(c). 
Notably, the higher HOMO-LUMO gap achieved from a large U value leads to the band gap in the solid, but the small HOMO-LUMO gap closes the band gap with band dispersion induced by the molecular periodicity in the solids.

This insulating electronic structure provides consistent behaviors to experimental observations.\cite{elsasser_power-law_2012}
First, the band gap $E_{gap}=56$~meV from DFT+GOU ($U=0.225$~eV) compares well with the experimentally measured one ($E_{gap}^{exp}=62$~meV), see Fig. 3(c).\cite{elsasser_power-law_2012}
Second, the frequency dependency of optical conductivity ($\sigma_1$) in the $\kappa$-ET-CN with DFT+GOU ($U=0.225$~eV) shows $\omega^{0.5}$ \& $\omega^2$ behavior under the independent particle approximation, see Fig. 3(d).\cite{elsasser_power-law_2012}
Because of the metallic band structure evaluated by DFT functional, its optical conductivity decreases with increasing light frequency.
On the other hand, the insulating band structure with DFT+GOU ($U=0.225$~eV) provides the $\omega^{0.5} +\omega^2$ behavior as experimental observation.\cite{elsasser_power-law_2012}
Third, the experimental pressure-induced metal-insulator transition is also achieved with DFT+GOU.\cite{kurosaki_mott_2005}
We evaluate the band gap of $\kappa$-ET-CN with respect to external pressure and U values for DFT+GOU, as shown in Fig.~3(e).
The higher U values that lead to the larger band gap at $P=0$~GPa provide the higher critical pressure for the metal-insulator transition.
When the experimental observation from resistance and NMR measurements indicates $P_c^{exp}\sim 0.37$~GPa critical pressure, the DFT+GOU with $U=0.225$~eV provides comparable behaviors ($P_c^{GOU}\sim 0.3$~GPa).
These results indicate that the electronic structure constructed by the accurately evaluated HOMO-LUMO gap of dimers provides a consistent description of the $\kappa$-ET-CN properties consistent material properties.

The superconducting dome induced by pressure in the $\kappa$-ET-CN originates from an aligned flat band state at the Fermi level.
From the DFT+GOU($U=0.225$~eV) calculation, we found that the gap of $\kappa$-ET-CN is opened under the ambient condition and is closed by the pressure near $0.3$~GPa, as shown in Fig.~3(e).
In the experiment, on the other hand, the hydrostatic pressure of $0.4$~GPa not only leads to the metal-insulator transition but also provides the highest T$_c$ value for the superconducting dome.\cite{kurosaki_mott_2005}
To understand this behavior, we investigate the electronic structure of $\kappa$-ET-CN with hydrostatic pressure with a value of the GOU of $0.225$~eV.
We found that $\kappa$-ET-CN has a flat band along the M-Y line at the Fermi level under $0.4$~GPa pressure, as shown in Fig. 4(a). 
This flat band induces a sharp peak in the density of state (DOS) at the Fermi level, as shown in the left panel of Fig. 4(b). 
Notably, this peak of DOS is unoccupied at lower pressure and is aligned at the Fermi level by high-pressure conditions, as shown in Fig. 4(b).
When we apply the higher hydrostatic pressure, the DOS at the Fermi level is reduced by the shifted downed flat band below the Fermi level, as shown in Fig. 4(c).
With this profile of DOS at the Fermi level, the critical temperature $T_c$ of superconductivity is evaluated using BCS $T_c$ equation as follows:\cite{bardeen_theory_1957} $k_B T_c=1.134~E_D~exp(-1/(N_F V))$, where $E_D$, $N_F$, and $V$ are Debye frequency, DOS at Fermi level and electron-phonon coupling potential, respectively.
We set the Debye frequency ($E_D=0.2$~eV) as the C-C strengthing vibration frequency of the BEDT-TTF molecule\cite{buzzi_photomolecular_2020} and the DOS at the Fermi level as its value from DFT+GOU calculation with $U=0.225$~eV. 
With these values obtained from our $ab~initio$ simulations, we evaluate the $T_c$ by varying the electron-phonon coupling potential ($V$) to find its appropriate value ($V=9.5\times 10^{-3}$DOS$^{-1}$) by comparing it to the experimental observations, as shown in Fig.~4(d).
Even though our approach misses the detail conditions, such as the pressure-dependent electron-phonon coupling potentials, Debye frequency, and U values, the BCS equation with the profile of DOS peak at the Fermi level with DFT+GOU($U=0.225$~eV) well reproduce the superconducting dome as the experimental observations.\cite{kurosaki_mott_2005}
This result reveals our description with DFT+GOU for the appropriate HOMO-LUMO gap ($\Delta$) of dimer systems in $\kappa$-ET-CN consistent behavior as experimental observation for the insulating nature and superconducting dome.
Notably, both DFT and DFT+U calculations, which do not succeed in increasing the HOMO-LUMO gap ($\Delta$) and opening the band gap ($E_{gap}$), fail to produce metal-insulator transition under pressure and a flat band alignment at the Fermi level (see SI, where we present the DFT and DFT+U band structures under various pressure conditions, and the DFT and DFT+U calculation results under pressure, which consistently provides a metallic band structure.).

The tight-binding Hamiltonian can be utilized to understand the fascinating many-body interactions in the $\kappa$-ET-X family.\cite{pratt_magnetic_2011,isono_quantum_2016,miksch_gapped_2021,suzuki_mott-driven_2022,Menke_Klett_Kanoda_Georges_Ferrero_Schäfer_2024,limelette_mott_2003}
Here, we evaluate the parameters for the tight-binding Hamiltonian of the $\kappa$-ET-CN electronic structure from the MO level of the dimer basis. 
For the simplest tight-binding model, we employ HOMO and LUMO states ($\phi_2^{\sigma}$) of a two-dimer system for the basis set (see Fig.~2(c)).
The hopping interactions between MO states, described by the $t$ term for the same sites (red or blue sites) in the next lattice and the $t'$ term for different sites (between red and blue sites), as depicted in Fig. 2 (e), introduce the band dispersion in the molecular solid. 
The rectangular lattice Hubbard Hamiltonian is given for fitting the DFT band structure as follows:
$H=\sum_{<ij>,\sigma}t(c^\dagger _{i,\sigma}c _{j,\sigma}+H.c.) + \sum_{[ij],\sigma}t'(c^\dagger _{i,\sigma}c _{j,\sigma}+H.c. ) +\Delta\sum_{i,\sigma}  (1-0.5n_{i,\sigma})$, when $<ij>$ and $[ij]$ are summations for the same sites in the nearest neighbor lattice and nearest neighboring different sites, respectively.
For the antiferromagnetic solution, we set site occupation (see Fig.~2(c)) as follows: $n_{1,\downarrow}=n_{2,\uparrow}=1$ and  $n_{1,\uparrow}=n_{2,\downarrow}=0$.
The metallic band structure with DFT calculation provides $t=50$~meV and $t'=39$~meV without the HOMO-LUMO gap ($\Delta=0$) of a charged dimer, which is consistent with the previous reports.\cite{kandpal_revision_2009,buzzi_phase_2021}
On the other hand, the insulating band structure from the DFT+GOU ($U=0.225$~eV) calculation provides $t=52$~meV, $t'=39$~meV, and $\Delta=239$~meV.
These parameters extracted from DFT and DFT+GOU indicate that GOU terms mainly modify each dimer state's HOMO-LUMO gap ($\Delta$) rather than hopping terms.
The band structure with $0.4$~GPa, which shows the metal-insulator transition, provides the following parameters: $t=56$~meV, $t'=45$~meV, and $\Delta=192$~meV.
It indicates the external pressure mainly reduced a dimer's HOMO-LUMO gap ($\Delta$) and increased inter-dimer hopping terms, which led the band gap close for $\kappa$-ET-CN.
These results indicate that our tight-binding Hamiltonian well reproduces the electronic structure evaluated by DFT+GOU and reinforces our central interpretation of the key role of a dimer's HOMO-LUMO gap for the electronic structure of the overall molecular crystal.
We suggest that these parameters can be exploited for the follow-up theoretical many-body studies for the spin liquid phase, light-enhanced superconductivity, and quantum criticality in  $\kappa$-ET-X family.\cite{isono_quantum_2016,pratt_magnetic_2011,buzzi_photomolecular_2020,miksch_gapped_2021,warawa_ultrafast_2023}
Notably, our tight-binding model includes the term for the HOMO-LUMO gap of the dimer ($\Delta$) to describe the insulating nature of $\kappa$-ET-CN. In contrast, previous models only consider the gap ($\Delta_1$) of the $\phi_1^\sigma$ states by estimating it from the hopping term between the nearest monomer ($\Delta_1 \sim 2t_1$ in Fig.~2(b)), not the HOMO-LUMO gap of a dimer ($\Delta$).\cite{kandpal_revision_2009,buzzi_phase_2021}
In the SI, we discuss details of tight-binding models with various basis set.\cite{coh_python_2022}

\section*{Discussion}
In summary, from the first principle calculation, we demonstrated the origin of insulating and superconducting dome phases in $\kappa$-ET-CN.
To understand the electronic structure of $\kappa$-ET-CN, we first investigated the MO level of (BEDT-TTF)$_2^{-1}$.
We found that hybrid functional provides the realistic HOMO-LUMO gap of the (BEDT-TTF)$_2^{-1}$, and its corresponding U ($\sim0.3$U$^{mol}_{ACBN0}$) value gives same HOMO-LUMO gap under the DFT+GOU calculation. 
When we extend the analysis on the bulk $\kappa$-ET-CN, the electronic structure of $\kappa$-ET-CN with DFT+GOU ($U=0.225$~eV) calculation provides four consistent behavior observed in experiments.
First, we reproduced the insulating electronic structure (E$_{gap}=53$~meV) as experimental observation (E$_{gap}^{exp}=62$~meV).
It indicates that the increased HOMO-LUMO gap in two (BEDT-TTF)$_2^{-1}$ dimers leads to the band gap in a solid system.
Consequently, the $\omega ^{0.5}+\omega ^{2.0}$ experimental behavior of the optical conductivity is well reproduced.
Third, pressure-induced metal-insulator transition at $P_{c}=0.3$~GPa is evaluated as experiment ($P_{c}^{exp}=0.37$~GPa).
Finally, the pressure-induced superconducting dome, which has a high DOS at the Fermi level due to a flat conduction band minimum state, is reproduced as experimentally observed. 
These results indicate that the MO energy levels of (BEDT-TTF)$_2^{-1}$ dimer are required for the electronic structure of the bulk $\kappa$-ET-CN, providing consistent properties from the first-principle calculation as experimental observations.
Additionally, we extracted the tight-binding parameters, which can be exploited in many body studies, revealing that pressure primarily influences the energy gap ($\Delta$) of the underlying dimer.
Our study provides a microscopic insight into the electronic structure of molecular solids, paving the way for a theoretical understanding of the exotic behaviors in the $\kappa$-ET family.\cite{isono_quantum_2016,pratt_magnetic_2011,buzzi_photomolecular_2020,miksch_gapped_2021,warawa_ultrafast_2023}

\section*{Methods}
\subsection{DFT calculation for solid}

We performed DFT calculations using the Quantum Espresso package.\cite{Giannozzi2017}
The wavefunction is described using the projector-augmented-wave method using pslibrary 1.0.0 version\cite{dal_corso_pseudopotentials_2014} and the plane wave basis set with an 80 Ry energy cut-off.
We employed the PBE-type exchange-correlation functional, hybrid functional, and DFT+GOU methods to describe the electron-electron exchange and correlations.\cite{Perdew1996,heyd_hybrid_2003,Shin2021}
The on-site Coulomb interaction for 8 states below the HOMO state and four states above the LUMO level is corrected by the DFT+GOU method.
The atomic geometry is fully relaxed with PBE plus van der Waals D3 (PBE+D3) functional.\cite{grimme_consistent_2010}
The calculated lattice constants for $\kappa$-ET-CN are $a=15.8$\AA, $b=8.46$\AA, $c=13.7$\AA, and $\beta = 110.8^\circ$ in the $P2_1/c$ space group under the ambient pressure.
The Brillouin zone is sampled with $4 \times 4 \times 4$ $\mathbf{k}$-point mesh for the molecular solid system.
The large cubic ($a=30.0$\AA) lattice in three-dimensional periodic boundary conditions is employed for the molecule calculations.
To achieve the pressured geometry of $\kappa$-ET-CN, we proceeded with lattice and ionic position relaxation with a given hydrostatic pressure along with a, b, and c axis using PBE+D3 functional.
The lattice relaxation is achieved with a given force criteria ($|F|<10^{-5} Ry/bohr$) for each atom and a given stress criteria ($|P|<0.005$~GPa).

The estimation of the effective Hubbard U for the $C-p$ and $S-p$ orbitals was done using the Octopus code,\cite{tancogne2020octopus} using the ACBN0 functional method.\cite{Agapito2015,tancogne-dejean_self-consistent_2017} We employed here a grid spacing of $0.3$ Bohr, a box made of the union of a sphere around each atom of the dimer, with a radius of 10 Bohr for each sphere, and pseudo-dojo PBE pseudopotentials.\cite{van2018pseudodojo} We employed the same relaxed geometry as in the rest of the simulations.

\subsection{Quantum chemistry calculation for molecules}

Molecular calculations on the BEDT-TTF dimer and two dimers were also performed using the Orca quantum chemistry software.\cite{neese_orca_2020}
The HOMO-LUMO gaps of both dimer and two dimers were carried out with the HF method, as well as with the DFT method by employing the generalized gradient approximation PBE exchange-correlation functional,\cite{Perdew1996} and the hybrid meta-generalized gradient approximation PW6B95 exchange-correlation functional with the def2-QZVPP.\cite{zhao_design_2005,weigend_balanced_2005}
We also double-checked our calculation with cc-pVDZ, and def2-SVP basis set.\cite{dunning_gaussian_1989,weigend_gaussian_2003}
Moreover, the DFT calculations with the PBE and PW6B95 functionals utilize the D3\cite{grimme_consistent_2010} and D4\cite{caldeweyher_generally_2019} dispersion corrections, respectively.

\section*{Data availability}
The data that support the findings of this study are available from the corresponding author upon request.

\section*{Code availability}
The code that supports the findings of this study is available from the corresponding author upon request.

\section*{References}

\providecommand{\noopsort}[1]{}\providecommand{\singleletter}[1]{#1}%

\section*{Acknowledgements}

We further acknowledge financial support from the European Research Council (ERC-2015-AdG-694097), the Cluster of Excellence "CUI: Advanced Imaging of Matter" of the Deutsche For-schungsgemeinschaft (DFG) - EXC 2056 - project ID 390715994, Grupos Consolidados (IT1249-19), and SFB925 "Light induced dynamics and control of correlated quantum systems". 
D.S. was supported by the National Research Foundation of Korea (NRF) grant funded by the Korea government (MSIT) (No. RS-2024-00333664 and RS-2023-00218180) and the Ministry of Science and ICT(No. 2022M3H4A1A04074153). This work was supported by the National Supercomputing Center with supercomputing resources including technical support (KSC-2024-CRE-0124). 

\section*{Author contributions}
D.S. and A.R. conceived the project. 
D.S., N.T., and F.P. performed the DFT and quantum chemistry calculations. 
D.S., N.T., and E.V.B. carried out the tight-binding model parameters.
D.S., M.B., and A.R. analyzed computed data and verified the experimental observations.
D.S., N.T., and A.R. wrote the manuscript with comments from all authors.

\section*{Competing interests}
The authors declare no competing interests.

\section*{Additional information}

\subsection{Supplementary information}
The online version contains supplementary information available at
https://doi.org/

\subsection{Correspondence}
Correspondence and requests for materials should be addressed to Dongbin Shin (email: dshin@gist.ac.kr), and Angel Rubio (email: angel.rubio@mpsd.mpg.de).

\clearpage

\begin{table*}[h]
    \centering
\caption{HOMO-LUMO gaps ($\Delta$) of (BEDT-TTF)$_2^{+1}$  and 2(BEDT-TTF)$_2^{+1}$ with various functionals}\label{tab1}%
\begin{tabular}{@{}llll@{}}
\toprule
\textrm{ HOMO-LUMO gap (eV)} & \textrm{PBE+D3}  & \textrm{PW6B95+D4} & \textrm{HF}\\
\midrule
(BEDT-TTF)$_2^{+1}$ ($\Delta$) &  0.14 & 1.3 &  2.96 \\
2(BEDT-TTF)$_2^{+1}$ ($\Delta$) &  0.11 & 0.66 &  2.39 \\
\end{tabular}
\end{table*}

\clearpage

\begin{figure*}[t]
    \centering
\includegraphics[width=0.95\textwidth]{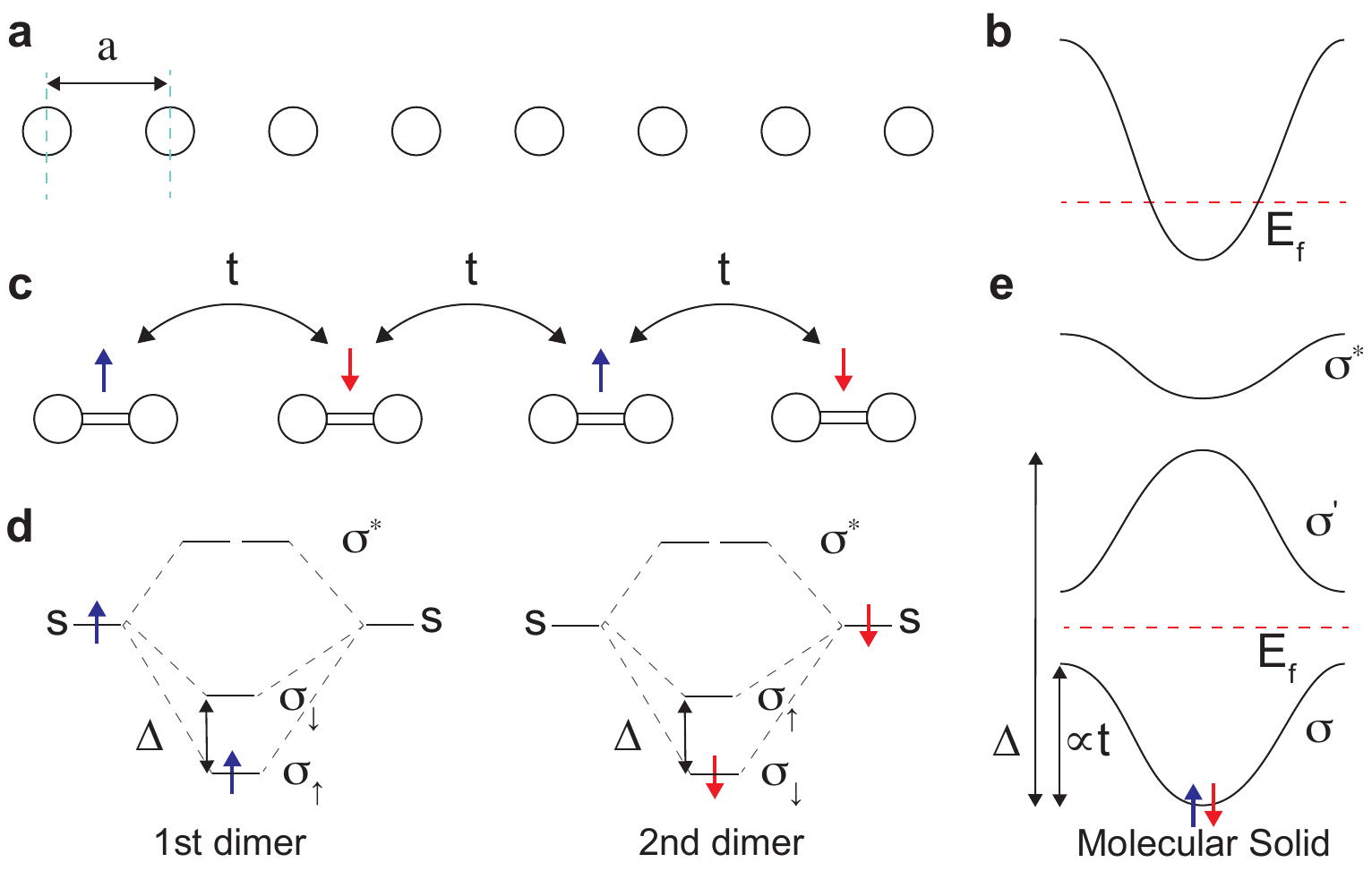}% Here is how to import EPS art
\caption{
    {\bf Electronic structure of quarter filled 1D hydrogen chain.}
\textbf{a} Atomic geometry of one-dimensional crystal with lattice parameter $a$. 
\textbf{b} The metallic band structure of a one-dimensional crystal in equal spacing with a quarter-filling. 
\textbf{c} One-dimensional crystal with periodicity $2a$ by dimerization with a quarter-filling.
\textbf{d} Schematic energy level of molecular orbital dimer level with on-site Coulomb interaction U.
\textbf{e} The insulating band structure of one-dimensional crystal by dimerization. 
}
\end{figure*}

\clearpage

\begin{figure*}[t]
    \centering
\includegraphics[width=0.7\textwidth]{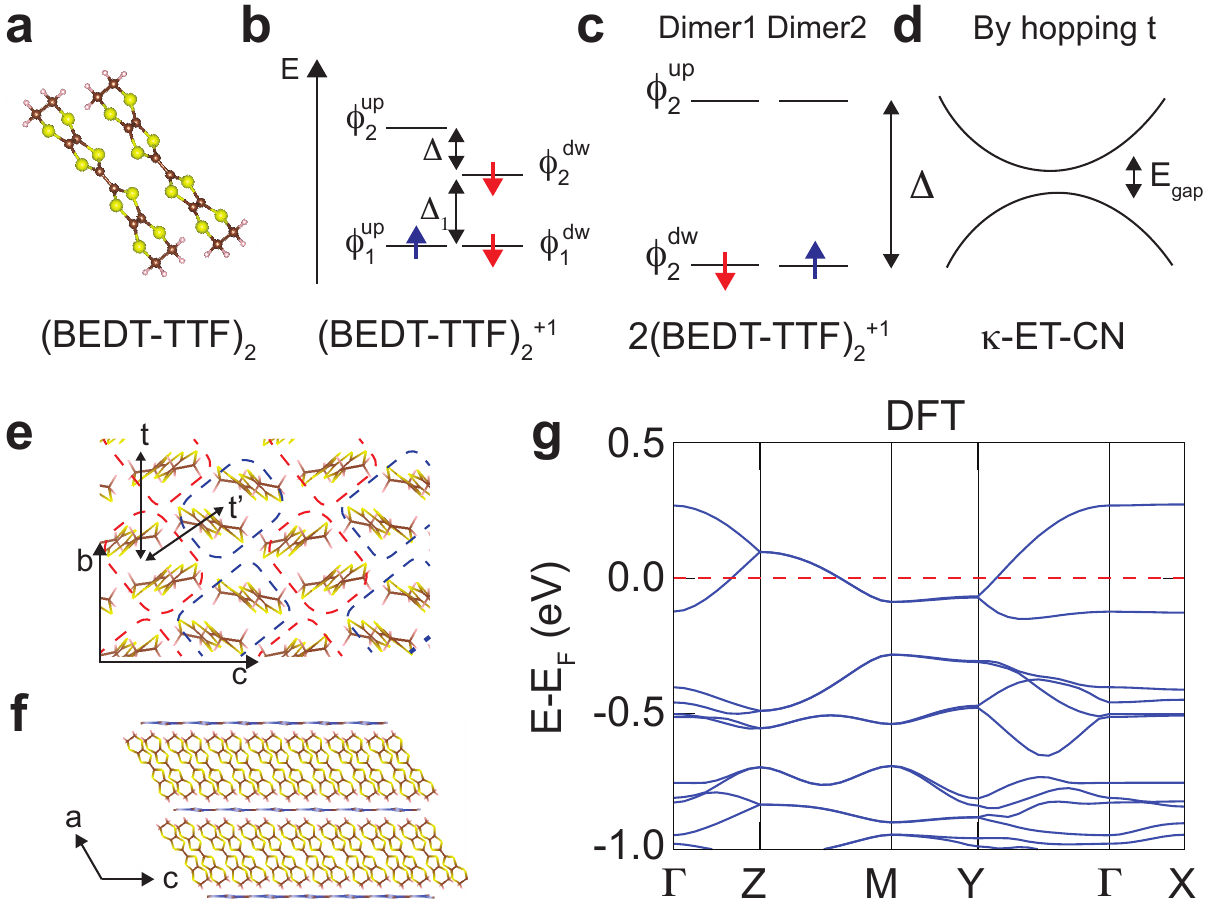}% Here is how to import EPS art
\caption{
    {\bf Atomic and electronic structures of $\kappa$-ET-CN from dimer.}
\textbf{a} Atomic geometry of BEDT-TTF dimer. 
\textbf{b-c} The schematic diagram for a molecular orbital energy level of (\textbf{b}) charged single dimer (BEDT-TTF)$_2^{+1}$, and (\textbf{c}) charged two dimer 2(BEDT-TTF)$_2^{+1}$.
\textbf{d} Schematic energy diagram of molecular solid constructed by 2(BEDT-TTF)$_2^{+1}$. 
\textbf{e-f} Atomic geometry of  $\kappa$-ET-CN in (\textbf{e}) a-c  and (\textbf{f}) b-c planes.
\textbf{g} Band structure of $\kappa$-ET-CN evaluated by DFT (PBE) functional.
In ($\textbf{e}$), the dashed blue and red boxes indicate the two-dimer basis for $\kappa$-ET-CN, and bold black double-sided arrows indicate the hopping between dimers for the nearest ($t$) and the next nearest ($t'$) neighboring dimer.
}
\end{figure*}

\clearpage

\begin{figure*}[t]
    \centering
    {\includegraphics[width=0.95\textwidth]{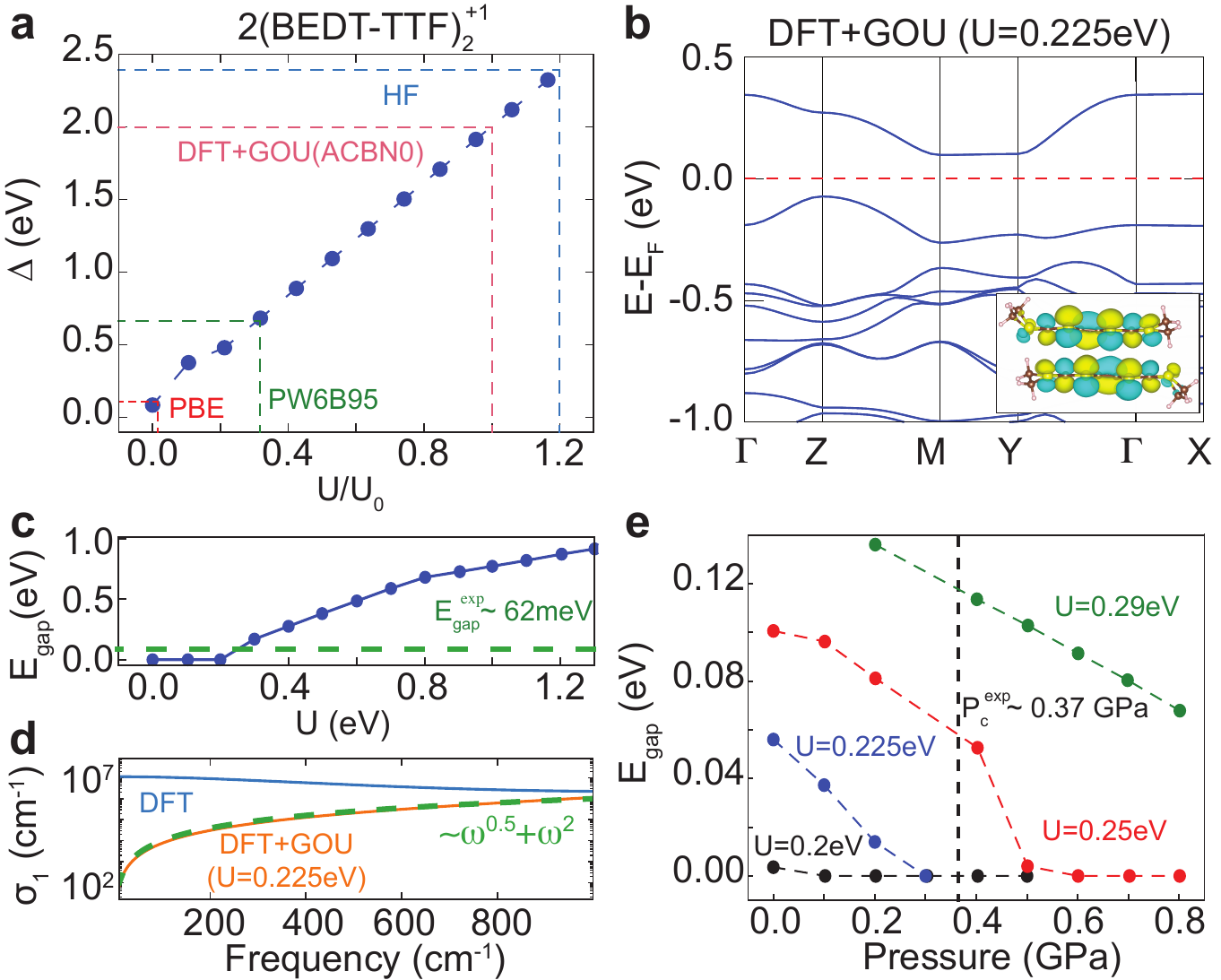}}% Here is how to import EPS art
  \caption{
    {\bf Insulating band structure of $\kappa$-ET-CN with the higher HOMO-LUMO gap of a dimer and reproduced properties.}
\textbf{a} The HOMO-LUMO gap of the charged two-dimer system computed with various functionals and U parameters for DFT+GOU.
\textbf{b} Band structure of $\kappa$-ET-CN evaluated by DFT+GOU method with $U=0.225$~eV.
\textbf{c} Band gap of $\kappa$-ET-CN calculated by various U parameters for DFT+GOU.
\textbf{d} Optical conductivity $\sigma_1$ of the $\kappa$-ET-CN evaluated by DFT and DFT+GOU($U=0.225$~eV). 
\textbf{e} U parameters- and pressure-dependency on the band gap of $\kappa$-ET-CN.
The inset of (\textbf{b}) indicates the $\phi_2^{\sigma}$ state of the dimer system, which consists of the valence band maximum and conduction band minimum states of $\kappa$-ET-CN.
}
\end{figure*}

\clearpage

\begin{figure*}
  \centering
{\includegraphics[width=0.95\textwidth]{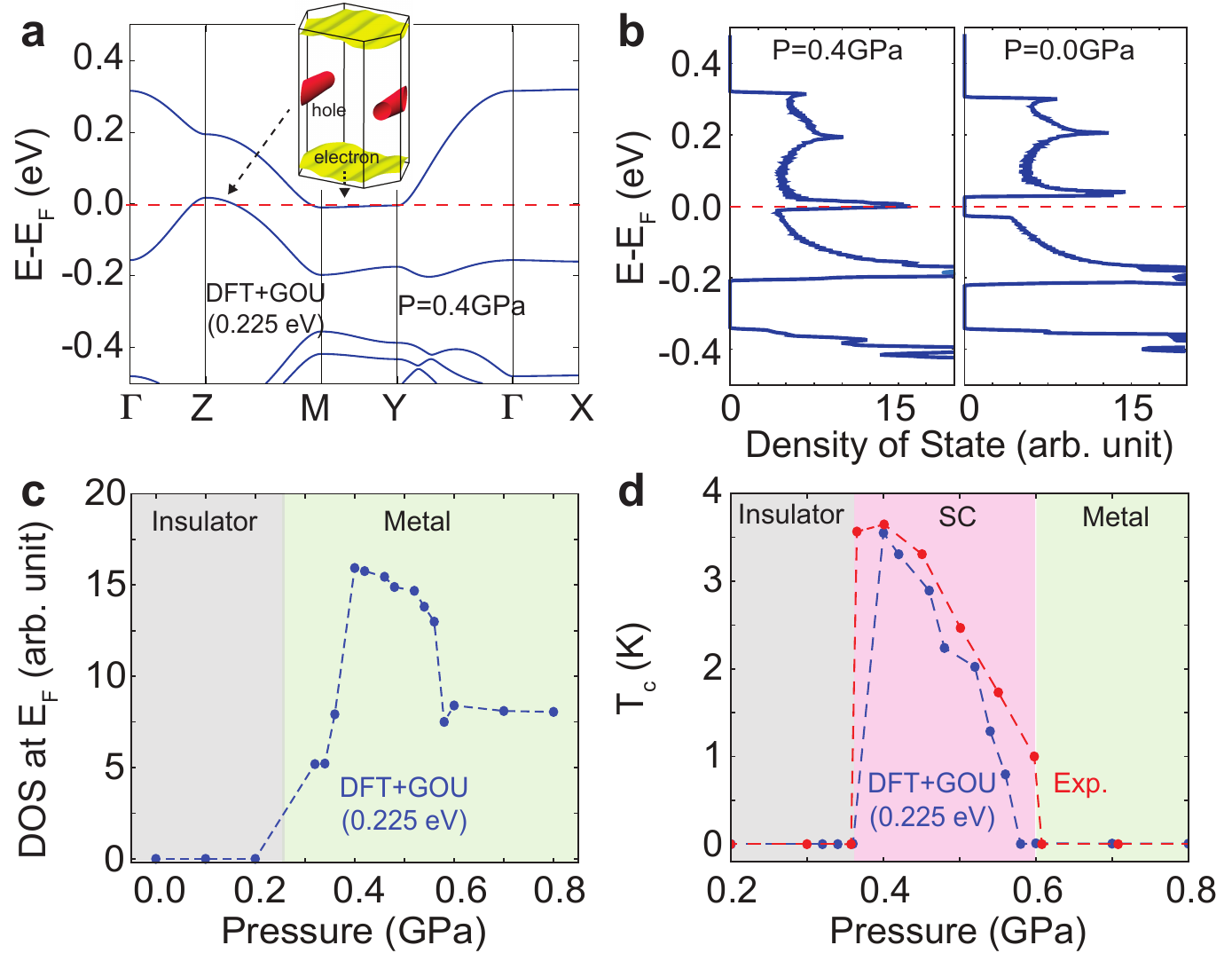}}% Here is how to import EPS art
\caption{
   {\bf Pressure-induced metal-insulator transition and superconducting dome.}
\textbf{a} Band structure of $\kappa$-ET-CN with $0.4$ GPa pressure using DFT+GOU ($U=0.225$~eV).
\textbf{b} Density of state of $\kappa$-ET-CN with (left) $0.4$ GPa, and (right) $0.0$~GPa pressures under the DFT+GOU ($U=0.225$~eV) calculation.
\textbf{c} Density of state at Fermi level with varying pressure using DFT+GOU ($U=0.225$~eV).
\textbf{d} Estimated critical temperature using BCS theory using DFT+GOU ($U=0.225$~eV) and $0.2$~eV for Debye cut-off energy and experimental data.\cite{kurosaki_mott_2005}
The inset of (\textbf{a}) indicates the Fermi surface of $\kappa$-ET-CN with $0.4$ GPa pressure using DFT+GOU ($U=0.225$~eV).
}
\end{figure*}

\end{document}